\documentstyle[twoside,fleqn,espcrc2]{article}

\newcommand{\LQ}{\Lambda_{QCD}}
\newcommand{\be}{\begin{equation}}
\newcommand{\ee}{\end{equation}}
\newcommand{\bea}{\begin{eqnarray}}
\newcommand{\eea}{\end{eqnarray}}

\title{QCD '98: status of the power corrections}

\author{V.I. Zakharov\address{Max-Planck Institut f\"ur Physik,
        80805 Munich, Germany.}}

\begin{document}

\begin{abstract}
We review status of the power corrections in QCD. The topics include
shape variables, unconventional $1/Q^2$ corrections, tachyonic gluon mass
as a fit parameter to the $1/Q^2$ corrections.
The selection of the material is determined mostly by results
presented at the Conference QCD'98 (Montpellier, July 1998). 
Some background comments are also included. 
\end{abstract}

\maketitle
\section{Introduction}

By power corrections to the parton model one understands terms
of order $(\LQ/Q)^n$ where the characteristic mass scale $Q$
is assumed to be much larger than $\LQ$. At first sight, the inclusion 
of the power correction in any consistent analysis is very problematic
since they are exponentially small in inverse running
coupling, $\sim exp(-n/2b_0
\alpha_s(Q^2))$, and, generally speaking, 
are subordinate to many orders in perturbative
expansion which makes them extremely sensitive to an exact definition
of $\alpha_s$ and so on. 

Thus, to develop a phenomenology one assumes in fact that 
the power corrections are enhanced
numerically and are responsible primarily for breaking of asymptotic freedom
at intermediate $Q^2$. The first signal for that came from the QCD sum rules
\cite{svz}. The central object here is the correlators of the currents $j$
with various quantum numbers:
\begin{equation}
\Pi_j (Q^2)~=~i\int exp(iqx)\langle 0|T\{j(x),j(0)\}|0\rangle ,\label{corr}
\end{equation}
($q^2\equiv -Q^2$)
and it is indeed known that the power corrections 
to the parton model predictions for $\Pi_j(Q^2)$
can be extracted with reasonable accuracy.

More relevant to the present discussion, the power corrections seem to 
be large in jet physics as well. In particular for a jet mass
one has \cite{barreiro}
\begin{equation}
\left({M_{jet}^2\over Q^2}\right)_{non-pert}~
\approx{2\langle k_{\perp}\rangle\over Q}.\label{bryan}
\end{equation}
where $Q$ is the jet energy and the intrinsic perpendicular
momentum of the partons is of order $\langle k_{\perp}\rangle\sim 0.5 GeV$
so that the nonperturbative correction is quite large. 

Most recently, the power corrections were discussed within the
renormalon framework (for reviews and further references see \cite{review}).
Renormalons provide with a derivation of equations like (\ref{bryan})
in a language directly related to the fundamental QCD.

Experimental and lattice data have been accumulated during 
last year and there have already been three talks at this 
Conference devoted entirely
to the power corrections \cite{fernandez,liuti,parrinello}.
I feel that the relevant theoretical framework was basicly reviewed 
during the previous conferences in the same series \cite{review}.
Thus we add only a few remarks, mostly on new theoretical works
which appeared meantime. Also, we include an overview of the 
original work \cite{cnz}.

\section{Shape variables.}

Data on $1/Q$ corrections to the thrust, C parameter and
heavy jet mass $M_h^2$ were obtained with high statistical
accuracy and presented at this Conference \cite{fernandez}.
The interest in these corrections, at least partly, was 
motivated by applications of the 
renormalon techniques  
\cite{webber,az,kg}.
In particular, the value of $n$ in $(\LQ/Q)^n$ is fixed reliably by
the renormalons for each variable and the presence of
the $\LQ/Q$ terms was demonstrated in case of the variables
mentioned above. However, 
to find  relations between various corrections
one needs still models because 
in principle any number of the renormalon chains gives power
corrections of the same order \cite{vaz,review}
and there is no way to evaluate all of them. 

Proceeding to the models, within the universality picture \cite{az,kg}
one keeps terms which contribute perturbatively  
in differential distributions of the variables 
as $\alpha_s^mln^kQ$ and continues such 
terms to the infrared region
(where, in general, they do not dominate any 
longer). As a result, all the $1/Q$ corrections
get expressed in terms of a universal factor
$E_{soft}=\int dk_{\perp}\gamma_{eik}
(\alpha_s(k^2_{perp}))$ where $\gamma_{eik}$ is the so called eikonal
anomalous dimension and the integral over the Landau pole is understood, 
say, as the principal value.
Within this model one gets, in particular, \cite{az}
\be
\langle 1-T\rangle_{1/Q}~=~{2\over 3\pi}\langle C\rangle_{1/Q}.
\label{safe}\ee
The data look as \cite{fernandez}
\bea
{1\over 2}\langle 1-T\rangle_{1/Q}={const\over Q}(0.511\pm 0.009)\\ 
{1\over 3\pi}\langle C\rangle_{1/Q} ={const\over Q}(0.482\pm 0.008),\label{exp}
\eea
and are in excellent agreement with the theory. 

There are further theoretical predictions. In particular \cite{az}:
\bea
\langle 1-T\rangle_{1/Q}=
\left({\langle M_h^2+M_l^2\rangle \over Q^2}\right)_{1/Q}\\
\left({\langle M_h^2\rangle\over Q^2}\right)_{1/Q}~\approx
\left({\langle M_l^2\rangle\over Q^2}\right)_{1/Q}\label{more}
\eea
Although the equation (\ref{more}) echoes the tube model (\ref{bryan})
its derivation within another renormalon-related approach 
\cite{webber} turned to be rather
painful. The point is that this is an approach based on a single
renormalon chain (or one-loop integration over the running coupling).
Such an approximation is tempting because of its simplicity
and various assumptions (like large $N_f$, naive non-abelianization,
freezing of the coupling...) were tried to justify it. However,
it is a basic fact about power corrections that the higher loops of 
the perturbative expansions, when projected onto the power corrections, 
all of them give contributions of the same order \cite{vaz,review}. It was
dramatically confirmed in case of the shape variables because in the one-loop
approximation \cite{webber} one of the jets is deprived of any non-perturbative
contributions and has mass strictly zero which cannot be true experimentally.

However, if one compares with experimental data, say, Eq. (\ref{safe})
then the both sides lose (for a two-jet event) a factor of 2 and 
the relation stays intact \cite{az}. Only once one proceeds, e.g., to relations (\ref{more}) the 
shortcomings of the model \cite{webber} become self-evident.
During the last year the so called Milan factor 
\cite{milan} was discovered within
the approach of Refs. \cite{webber}. Which is, in essence, the observation
that two loops contribute to the shape variables no less than one loop.
The explicit form of the Milan factor emerges from a careful analysis 
of the infrared radiation on the two-loop level.
As a result, in cases when same variables were considered within both
approaches \cite{az} and \cite{webber}
\be
(DW)\times (Milan~factor)~=~(universality)
\ee
where $(DW)$ stands for the original predictions of the model \cite{webber}.
It might worth noting that the Milan factor does not necessarily 
keeps exactly the same contributions which are absorbed into 
the factor $E_{soft}$ (see above). However, the both models 
share the property 
that for a two-jet event nonperturbative 
contributions are the same for each of the jets. And the both derivations 
are still models, not a proof. 

Experimentally,
\be
2 \left({\langle M_h^2\rangle\over Q^2}\right)_{1/Q}~=~{const\over Q}(0.616\pm0.018)
\ee
where the constant is the same as in Eqs. (4,5).
kl
As is emphasized in \cite{fernandez} the experimental data 
on the jet broadening parameters, $B_T,B_W$ contradict
the Milan factor (these variables have not been treated within the
universality approach so far). On the other hand, there
is no contradiction with hadronization models. My superficial
judgment would be that the Milan factor will yield
to the hadronization models as a result of further analysis. The reason is that
the renormalons do incorporate the tube model as a hadronization model
\cite{az}.

\section{Power corrections in DIS.}

Theory of the power corrections in deep inelastic scattering has 
been developed so far only on the one-loop level (i.e., a single renormalon
chain) \cite{stein}. Which might be not such a bad approximation 
(unlike the case of the $e^+e^-$ annihilation, see above)
since there is a single quark which radiates.
And indeed first calculations \cite{gupta} with inclusion of the Milan factor
 confirm these expectations. 
However, even on the one-loop level the Landau-pole 
parametrization of the power corrections produces 
in case of the $F_L$ structure function a result which 
is in variance with a
single renormalon chain predictions \cite{az5}. 
 Thus, the theoretical predictions appear to depend on details
of the infrared cut off, the possibility always looming over the
renormalon-based approaches. Also, the extraction of the numerical 
value for the power corrections in DIS at least in some cases
depends crucially on the subtraction of the perturvative pieces
\cite{kataev}. 

A new and systematic extraction of the power correction to the non-singlet
structure function $F_2$ from the experimental data was presented
at this Conference \cite{liuti}. The power corrections were studied 
as a function of $n$ where $n$ is the number of the moment from
the structure function $F_2(x)$ and the $n$-dependence appears to agree
with the renormalon-based predictions.

\section{Power corrections beyond the OPE.}

So far we discussed power corrections associated with infrared region.
Although the theoretical predictions may vary in this case
as well, the  physical picture behind the IR corrections is 
simply the growth of the
running coupling at large distances and this mechanism seems safe.
 
During the last couple of years unconventional $1/Q^2$ corrections
which go beyond this simple picture have also been 
discussed.  At this Conference results on measuring $1/Q^2$
terms in $\alpha_s$ have been presented \cite{parrinello}.  Papers in
Ref. \cite{parrinello} contain not only results of the lattice
measurements but an exposure of the theoretical framework as well and
we would not like to overlap with this discussion.  The $1/Q^2$
corrections are viewed in Ref. \cite{parrinello} as arising within the
dispersive approach, or from removal of the Landau from the running
coupling \cite{grunberg,az4}. In fact, there is another way to
introduce the $1/Q^2$ corrections which go beyond the standard OPE,
that is via the UV renormalons \cite{yamawaki}. 
It has not been clarified yet, whether there is any
connection between the two ways of viewing the unconventional $1/Q^2$
corrections.

The leading ultraviolet renormalon brings perturbative expansions of
the form: \be \left(\sum_na_n\alpha_s^n(Q^2)\right)_{UV}~=~ \sum_n
n!(-b_0)^n\alpha_s^n(Q^2)
\label{uv}.\ee 
Because of the $n!$ divergence of the expansion coefficients the sum
(\ref{uv}) is not defined. It is quite common to assume however that
the Borel summation is the correct recipe to deal with the
divergence. This hypothesis amounts to replacing the growing branch of
the perturbative expansion (\ref{uv}) by its integral representation:
\be \sum_{N_{cr}}^\infty n!(-b_0)^n\alpha_s^n \rightarrow \int
{(\alpha_sb_0 t)^{N_{cr}}exp(-t)dt\over 1+\alpha_sb_0t}\label{renorm}
\ee where $N_{cr}=1/b_0\alpha_s$ is the value of $n$ for which the
absolute value of the terms in the series reaches its minimum and the
right-hand side is readily seen to be of order ${\LQ^2\over Q^2}$: \be
{1\over 2} \left(a_n\alpha_s^n(Q^2)\right)_{n=N_{cr}}\sim {\LQ^2\over
Q^2}.\ee It is amusing to observe that this correction comes from huge
virtual momenta of order $p^2\sim Q^2\cdot exp(N_{cr})$.

Thus, if the theory is defined with an intrinsic UV cut off, like
the lattice theory, then the contribution of the UV renormalon may
well be irrelevant.  Then an alternative language
of dispersive approach to the running coupling 
\cite{grunberg,az4} can be utilized. Within this approach,
one may think in terms of the removal
of the Landau pole from the coupling: 
\be {1\over
lnQ^2/\LQ^2}~\rightarrow{1\over lnQ^2/\LQ^2} -{\LQ^2\over (Q^2-\LQ^2)}
\label{removal}.\ee
To justify this modification of the coupling one argues that to any
finite order in perturbation theory the coupling satisfies dispersion
relations with cuts at physical $s>0$.  The procedure, which has been
discussed since the fifties, clearly introduces a $\LQ^2/Q^2$
correction to $\alpha_s$ at large $Q^2$.

The bulk of the discussion is build on the belief that the replacement
(\ref{removal}) results in a new universally defined coupling which
can be used then for phenomenological purposes.
However, it was argued in Ref. \cite{az4}, and I still consider the 
argument convincing, that
the dispersive approach does not end up with a universal redefinition of
the coupling. 
Instead,
all the terms in $\alpha_s(Q^2)$ perturbative expansion collapse to
the same order power correction once the Landau pole is removed
from the dispersive representation for $\alpha_s^n(Q^2)$.
In more detail \cite{vz2}: 
\be
\sum_{n}a_n\alpha_s^n(Q^2)\rightarrow \sum_{n}a_n\alpha_s^n(Q^2)-\sum_n
{a_n\over n!b_0^n}{\LQ^2\over Q^2}\label{borel}
.\ee

Thus to establish a connection (if any) 
of the removal of the pole from dispersion representations
with the UV renormalons one has to consider the whole series 
in $\alpha_s(Q^2)$. In particular, let us apply 
(\ref{borel}) to the case $a_n=n!(-b_0)^n$. Then 
the resulting power correction,
$\sum_n (-1)^n\LQ^2/Q^2$, is still poorly defined. If, however, we invoke the
Borel summation,
\be
\left(\sum_n(-1)^n{\LQ^2\over Q^2}\right)_{Borel}~=~{1\over 2}{\LQ^2\over Q^2}
,\ee 
then the power corrections resulting from 
the procedures (\ref{borel}) and (\ref{renorm})
are the same. This example demonstrates that there could be a close connection
between the Borel summation of the UV-renormalon series (\ref{renorm})
and the removal of the Landau pole (\ref{removal}) from dispersion relations.
Moreover, one may argue that if introduction of the UV cut off into the
theory removes the UV renormalon, the coupling should be redefined
along the lines of Eq. (\ref{removal}) so that the theory with and without the 
UV cut off remains the same 
at relatively low momenta on the level of the $1/Q^2$ terms. 
Thus, one is invited to generalize the logic of the renormalization group
to include the $1/Q^2$ corrections as well. 

Let us now turn to another issue, namely the physical meaning
of the replacement (\ref{removal}). To this end we need a more precise
definition of $\alpha_s(Q^2)$
and we assume now that $\alpha_s(Q^2)$
is defined in terms of the Fourier transform of
the potential of heavy quarks $V(r)$ \cite{az3}.
It is obvious then that the 
change (\ref{removal})
results in a linear correction to the potential at {\it short} distances
$r\ll \LQ^{-1}$:
\be
\lim_{r\rightarrow 0}V(r)=-{4\alpha_s(r)\over 3r}+(const)r\label{pot}
.\ee
On the other hand, within the so to say standard QCD the leading 
power correction at {\it short} distances is of order $r^2$. This 
conclusion is 
based solely on the assumption that the nonperturbative 
fluctuations in QCD are of large
scale, $\sim \LQ^{-1}$ (for references and further explanations see
\cite{az3}). Thus, introduction of the new corrections (\ref{removal})
assumes small-size nonperturbative fields
and a particular picture of the vacuum properties which results in
this effect is described in Ref \cite{az3}.
Arguments that at least for some definitions of $\alpha_s$
the $1/Q^2$ piece can be associated only with
short distances were also given most recently in Ref. \cite{kivel}.

\section{Tachyonic gluon mass.}

 The $1/Q^2$ corrections discussed in the previous section go beyond
the standard OPE. Detection of the new type corrections
through phenomenology 
would be of great interest. In this section we will discuss
phenomenology in terms of a tachyonic gluon mass which is assumed to mimic
the short-distance nonperturbative effects  \cite{cnz}.

First, let us note that not all the $1/Q^2$ corrections in QCD are
associated with short distances. For example, in case of 
DIS the $1/Q^2$ corrections are coming from the 
IR region and perfectly consistent with the OPE.
Thus, the class of theoretical objects for which an observation
of the $1/Q^2$ corrections would signify going beyond the OPE is
limited. One example was the potential $V(r)$ discussed above.
Other examples are the correlator functions (\ref{corr}) where
the IR-sensitive power corrections start with $Q^{-4}$ terms \cite{svz}.

Thus, we concentrate on this set of variables and  
an interesting question is whether
there is room for introduction of sizable non-standard $1/Q^2$ 
corrections. The answer seems to be in positive
\cite{cnz}.
First of all, the potential 
(\ref{pot}) itself has been measured on the lattice
down to  fairly
short distances \cite{bali} and there is no sign of 
the change from the linear term
$kr$ (where $k$ is the string tension) to a quadratic correction 
$cr^2$ as is predicted by the OPE.
Thus at this time one is free to speculate that the form
(\ref{pot}) continues to $r\rightarrow 0$ with $const~\sim~k\approx
0.2GeV^2 $.

Even if one accepts this assumption, it is far from being trivial
to relate the constant $k$ to the scale of the 
$1/Q^2$ corrections to other quantities.  
Qualitatively, however, one may hope that
introduction of a tachyonic gluon mass at short distances  
would imitate the effect of the 
$\LQ^2/Q^2$ corrections. Indeed, the linear term in (\ref{pot})
and with $const\sim k$ can be imitated \cite{vz2} by the Yukawa 
potential with a gluon mass $\lambda $:
\be
{4\alpha_s\over 6}\lambda^2~\sim~-~k\label{mass}
.\ee
This picture with a tachyonic gluon mass can be consistently used at one-loop
level as well. 

Of course Eq. (\ref{mass}) may serve only for a rough estimate.
First of all, there are no error bars on the value of $k$ at short
distances since the measurements \cite{bali} were not dedicated to
the short distances. 
Moreover, Eq. (\ref{mass}) assumes that the 
short-distance potential is due to
a vector-like exchange while at large distances the $kr$ term corresponds 
to a scalar exchange and there is no evidence for a change \cite{bali}.
However, the reservations (which are numerous) should 
not mask the fact that the gluon mass is large according to 
(\ref{mass}). And the real question is
\cite{cnz} whether a kind of large tachyonic gluon mass is
admitable in view of the known properties of 
$\Pi_j (Q^2)$.

One of the basic quantities to be determined from the theory is 
the scale at which
the parton model for the correlators (\ref{corr})
gets violated considerably via the power corrections. Technically,
one studies usually $\Pi (M^2)$ where \cite{svz}
\be
\Pi_j (M^2)~\equiv~{ Q^{2n}\over (n-1)!}\left({-d\over dQ^2}\right)^n\Pi_j (Q^2)
\ee
in the limit where both $Q^2$ and $n$ tend to infinity so that their
ratio $M^2\equiv Q^2/n$ remains finite. Moreover, within the standard OPE
the correlators $\Pi (M^2)$ at large $M^2$ is represented as:
\bea
\Pi_j(M^2)\approx (parton~model)\cdot\\
\left(1+{a_j\over lnM^2/\LQ^2}+{b_j\over M^4}+O((lnM^2)^{-2}M^{-6})\right)
\eea
where the constants $a_j,b_j$ depend on the channel, i.e. on the quantum
numbers of the current $j$. The terms of order $1/lnM^2$ and $M^{-4}$
are associated with the first perturbative correction and the gluon
condensate, respectively. To characterize the scale of the power
corrections, one may introduce \cite{novikov} the notion of
$M^2_{crit}$ which is defined as the value of $M^2$ at which 
the power corrections become, say, 10\% from the unit. The meaning of 
$M^2_{crit}$ is that at lower $M^2$ the power corrections blow up.

In the $\rho$-channel, 
\be
M^2_{crit}(\rho-channel)~\sim~0.6~GeV^2\label{normal}
\ee
which is determined by the value of the gluon condensate
$<\alpha_s(G_{\mu\nu}^a)^2>$ and agrees well with independent 
evaluation of $M^2_{crit}$ from the experimental data
on the $e^+e^-$ annihilation \cite{svz}. 
In the language relevant to the present review,
the gluon condensate represents IR renormalons. 

If one proceeds to other channels, in particular to the 
$\pi$-channel and to the $0^{\pm}$-gluonium channels, nothing special
happens to $M^2_{crit}$ associated with the IR renormalons.
However, it was determined from independent arguments that the
actual values of $M^2_{crit}$ do vary considerably in these
channels \cite{novikov}:
\bea 
M^2_{crit}(\pi-channel)~\ge~ 1.8~GeV^2\\\label{pion}
M^2_{crit}(0^{\pm}-gluonium~channel)~\ge~ 15~GeV^2\label{gluonium}
.\eea
These lower bounds on $M^2_{crit}$ are obtained from the values of
$f_{\pi}$ and of the quark masses in the pion channel, and from a low-energy 
theorem in the gluonic channel. Such values of $M^2_{crit}$ 
cannot be recoinciled with the IR renormalons.

Now, a new term proportional to $\lambda^2$ is added
to the theoretical side of $\Pi_j(M^2)$ which becomes: 
\be
\Pi(M^2)\approx (parton~model)(1+{a_j\over lnM^2}+{b_j\over M^4}+{c_j\over M^2})
\ee
where $c_j$ is calculable in terms of $\lambda^2$ \cite{cnz}:
\be
c_{\pi}\approx 4c_{\rho}={4\alpha_s\over 3\pi}c_{gluonium}=
{4\alpha_s\over \pi}\lambda^2.
\ee
Phenomenologically,
in the $\rho$-channel there are severe restrictions \cite{narison}
on the new term $c_j/M^2$:
\be
c_{\rho}~\approx~-~(0.03-.07)~GeV^2\label{constr}
.\ee
Remarkably enough, the sign of $c_{\rho}$
does correspond to a tachyonic gluon mass
(if we interpret $c_{\rho}$ this way).
Moreover, when interpreted in terms of $\lambda^2$ the constraint (\ref
{constr}) does allow for a large $\lambda^2$, say, $\lambda^2=-0.5GeV^2$.

As for for the $\pi$-channel one finds now a new value of $M^2_{crit}$
associated with $\lambda^2\neq 0$:
\be
M^2_{crit}(\pi-channel)\approx~4\cdot M^2_{crit}(\rho-channel)\ee
which fits nicely the Eqs.
(\ref{normal}) and (22) above. 
Moreover, the sign of the correction in the $\pi$-channel 
is what is needed for phenomenology \cite{novikov}.
Fixing the value of $c_{\pi}$ to bring the theoretical $\Pi_{\pi}(M^2)$
into agreement with the phenomenological input one gets
\be
\lambda^2~\approx~-0.5~ GeV^2
.\ee

Finally, we can determine the new value of $M_{crit}^2$ in the 
scalar-gluonium channel and it turns to be what is 
needed for the phenomenology,
see Eq (\ref{gluonium}).
Thus, qualitatively the phenomenology with a tachyonic gluon mass
which is quite large numerically stands well to a few highly nontrivial
tests.

It is worth emphasizing that the $\lambda^2$ terms represent 
nonperturbative physics and limit in this sense the range of applicability
of pure perturbative calculations. 
This nonperturbative piece may well be 
much larger than some of the perturbative corrections which are 
calculable and calculated nowadays.

Further crucial tests of the model with the tachyonic gluon mass could be 
furnished with measuremants of various correlators $\Pi_j(M^2)$
on the lattice.

\section{Conclusions}

We discussed briefly the status of the power corrections
both from IR and UV regions. The infrared renormalons
produce a picture close to the hadronization
models. Phenomenologically, there is room for a new 
$\LQ^2/Q^2$ correction coming from the ultraviolet.
If confirmed, such a correction would be of great interest. 

I would like to acknowledge gratefully the 
hospitality of the Centre National de la
Recherche Scientifique (CNRS), and especially of S. Narison, 
during the stay at the 
University of Montpellier where part of this
work was done. I am also grateful to R. Akhoury, G. Bali, K.G. Chetyrkin,
and S. Narison for interesting discussions and remarks.

\end{document}